# Rapid Measurement of Spectral Characteristics by Correlation Matching Method

Chol-Sun Kim , Chol-Su Kim and Song-Jin Im

Department of Physics, **Kim Il Sung** University, Ryongnam Dong, Taesong District, Pyongyang, DPR Korea

Abstract

In this paper, we have established the couple system of a spectroscope, CCD and computer and proposed a method of the rapid measurement on spectral characteristics such as central wavelengths, relative intensities, sensitivity lines and the wavelength range and image pixel of the spectral images of a material by using the correlation matching method for the image discernment of digital spectra.

We study on the rapid identification of spectroscopic spectra and the determination of its characteristics by the correlation matching of digital images. It is advantageous to apply image matching [1] and correlation[2,3] rather than picture comparison method[4,5] on the recognition and identification of digital images obtained by means of image input devices such as CCD [6-12].

In this paper, we have proposed a method for rapid determination of the elements and their spectral wavelength, the dimensions of and image and wavelength region by applying correlation matching on the image recognition of digital spectroscopic spectra.

### 1. Correlation Computation of Digital Spectral Image

A spectral signal like the one from a gas-discharge lamp which emits from an atom in a gas-state with a high temperature of a few thousand degrees contains characteristics[12-14] like wavelength ($\lambda$), intensity(I) and such as shown in expression (1).

$$I_{ik} = A_{ik} \cdot \frac{hc}{\lambda_{ik}} \cdot N_0 \cdot \frac{g_i}{g_0} \cdot e^{-\frac{E_i}{KT}} \qquad (1)$$

In expression (1) $I_{ik}$, $\lambda_{ik}$ are the intensity and wavelength of the spectral line. $N_0$ is the total number of atoms of the given element. $g_0$ and $g_i$ are the statistical weight of the atom on the ground level and excitation level, respectively. k is the Boltzmann constant. T is the absolute temperature of the light source. $E_i$ is the excitation energy of the atom, $A_{ik}$ is the transition probability. h and c are constants.

The specter of atoms emitting in a gas-state are inherent[15] and the spectral lines of a given element have similarities and each other correlation[1-3,16], even if they are obtained at different times from various materials.

When we investigate the similarity in the intersecting area between the two spectroscopic specters T(j,k) and F(j,k) which have been recorded on a computer as digital data in the type of a N×M array by means of a CCD, the index of similarity is determined by the subtraction of the luminosities of each pixel on the two images.

Let's define F(j,k) as the standard digital spectral image and T(j,k) as the probe digital spectral image. The similarity index D(m,n) of two pixels m,n on the probe image is as follows.

$$D(m,n) = \sum_j \sum_k [F(j,k) - T(j-m, k-n)]^2 \qquad (2)$$

Expression (2) can be transformed as follows.

$$D(m,n) = \sum_j \sum_k [F(j,k)]^2 - 2\sum_j \sum_k F(j,k) \cdot T(j-m, k-n) + \\ + \sum_j \sum_k [T(j-m, k-n)]^2 = D_1(m,n) - 2D_2(m,n) + D_3(m,n) \qquad (3)$$

Only $D_2(m, n)$ confirms the correlation between the standard image and probe spectral image. At the position where the two spectral images match best, this correlation value becomes maximum and the luminosity variance between the two images becomes minimum.

If we define the standard iron spectral image of spectroscopic specter as F(j, k) and the probe iron spectral image as T(j, k), the two spectral images are similar and the cross correlation coefficient[1] is as follows.

$$C(m,n) = \frac{\sum_{j,k} [F(j,k) - \overline{F}_{m,n})][T(j-m, k-n) - \overline{T}]}{\sqrt{\sum_{j,k} [F(j,k) - \overline{F}_{m,n}]^2 \sum_{j,k} [T(j-m, k-n) - \overline{T}]^2}} \qquad (4)$$

$\overline{F}_{m,n}$ is the average value of the standard spectral area which overlaps with the probe spectral image when the upper right-hand corner of the probe spectral image is conformed with point (m, n) of the standard spectral image, $\overline{T}$ is the average value of the probe spectral image.

The cross correlation coefficient has a value between -1 and 1, and if this value becomes maximum, the two images are most similar. Thus, if we find the position of this value, we can distinguish to which part of the standard spectral image the probe spectral image corresponds. If we calculate according to expression (4), the cross correlation coefficient is given in a matrix format. But in case of a spectroscopic analysis spectral image, the longitudinal variance has nearly no significance. If we average along the longitudinal axis, the cross correlation becomes a vector and the computation amount reduces enormously.

We can find the wavelength of the spectral lines in question using the data of wavelength value corresponding to every pixel of the standard specter.

**2. Device and Method for Experiment**

The experimental setting[6,8,17-19] for the recording of the spectroscopic spectral image and its result processing is shown in Fig. 1.

AC arc generator "ДГ-2" has an arc current of 4-6A and an electrode gap of 2-3mm.

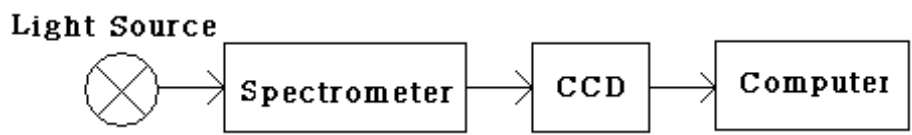

Fig 1. Setting for the recording and processing of spectral images

The crystal spectrometer[4,5] "ИСП-28" has a slit width of $10\mu m$ and a scatter band of 200-700nm and its output is coupled with a CCD camera which has a resolution of 640×480 pixels, a transmission rate of 30frame/s and a spectral reception band of 350-700nm.

We have made a database of the symbol of element, wavelength, relative intensity and sensitivity line of 337 spectral lines, including 136 sensitivity lines of 56 elements like Li, Be, C, Na, Mg, Al, Si, K, Ca, Ti, V, Cr, Mn, Co, Ni, Cu, Zn, Mo, Ag, Cd, Sn, W, Pt, P, S, Hg, Pb from the spectral area of 350-700nm. And we have inserted this Fe standard spectral database into a computer.

The dispersion curve drawn by cubic spline interpolation of the pixel values corresponding to the wavelength values in the spectral area of 350-700nm is shown in Fig 2.

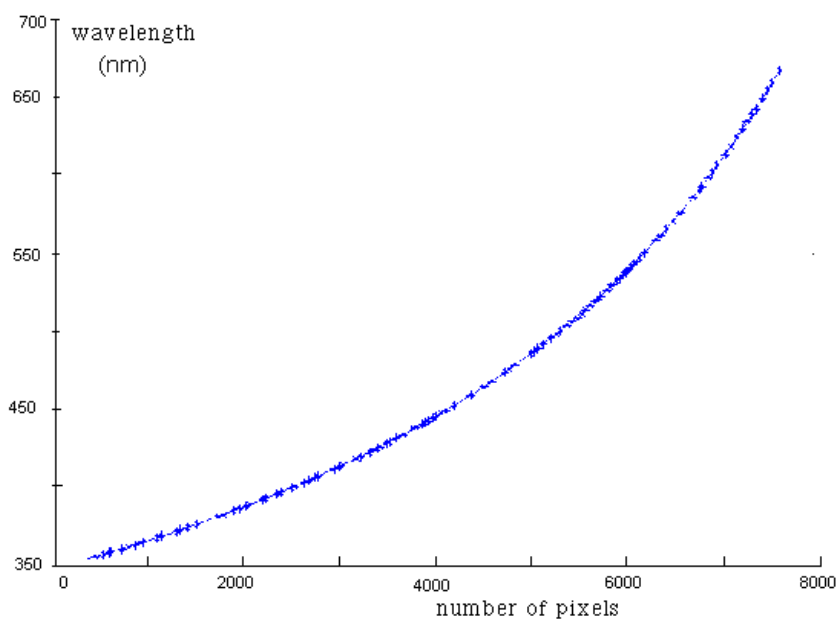

Fig2. Dispersion curve

The spectral images recorded by the CCD camera is processed with the program explained in Fig 3, thus giving output of the computed wavelength, element type, relative intensity, sensitivity lines, image size and wavelength area.

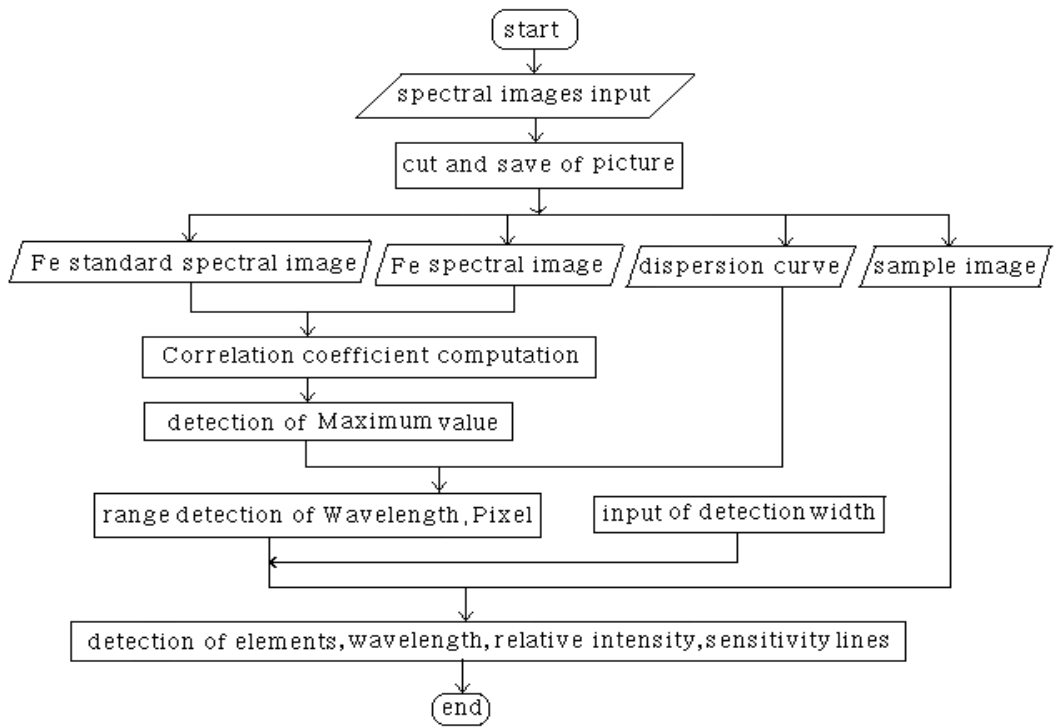

Fig3. Algorithm for spectral characteristics measurement

The program[20] is built with Matlab 7.4.0 and takes into consideration the image matching method [1]. Image matching method is a method to compare and compute the similarity between a model image and the intersecting part of the given image. The similarity index is a value based on the intensity variance between each pixel of the model image and given image.

### 3. Experimental Results and Analysis

The measurement results of spectral lines[21] 568.27nm, 568.82nm, 588.996nm, 589.59nm emitted from a Na light source are given in table 1 and Fig 4.

Table 1. Image Matching Characteristics

| Wavelength range(nm) | Pixel range(px) | Maximum position value | Correlation coefficient value |
|---|---|---|---|
| 553.141−599.898 | 6206−6845 | 6845 | 0.4893 |

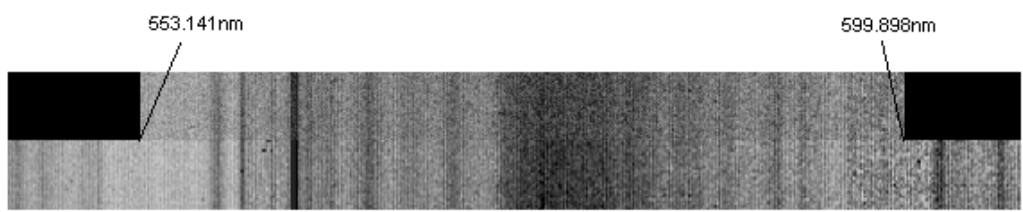

Fig 4. Image matching by cross correlation coefficient

The upper part of the figure shows the sample iron specter and the lower part shows the standard iron specter.

The correlation coefficient helps find the pixel position of maximum value, fulfill image matching and obtain the pixel and wavelength area of the sample image.

The specter from Na light source (upper part) and standard iron specter (lower part) are shown in Fig 5. The measurement results from positions 1, 2, 3 and 4 are given in table 2, where the measured lines are sensitivity lines. The difference of wavelengths of the emission lines from the Na light source and the wavelengths at the selected positions are 0.01-0.06nm, thus being in the range of detection width(0.1nm).

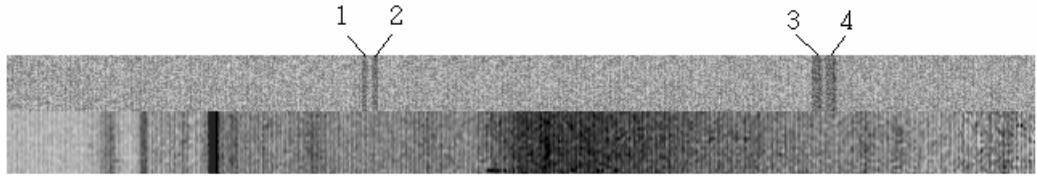

Fig 5. Spectral image of Na in the wavelength range of 553nm-600nm

Table 2. Measurement results (detection width = 0.1nm)

| No | element | wavelength(nm) | position(nm) | Sensitivity line |
|---|---|---|---|---|
| 1 | Na | 568.270 | 568.280 | ○ |
| 2 | Na | 568.820 | 568.790 | ○ |
| 3 | Na | 588.996 | 589.654 | ○ |
| 4 | Na | 589.590 | 589.552 | ○ |

Similarly, we have used Pb II 560.89nm sensitivity line[21] in the separation of yellow Pb series(Pb content 0.7-2%) from copper alloys. We have selected Fe as auxiliary electrode (diameter 8mm, length 69mm) and implemented image matching as shown in table 1 and Fig 4 and given results in Fig 6 and Fig 7.

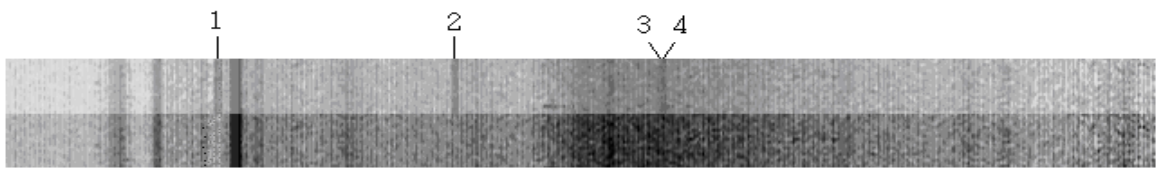

Fig 6. Sample image **with Pb**(upper part) and iron spectral image(lower part) between 553-600nm

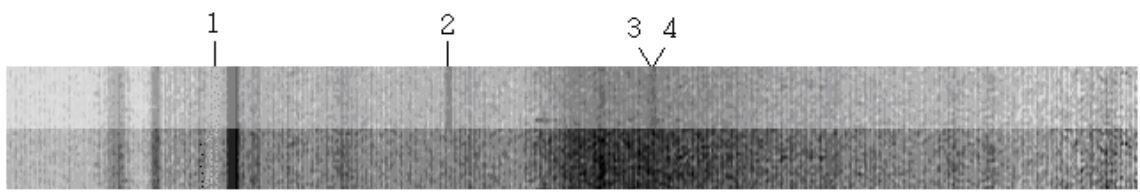

Fig 7. Sample image **without Pb**(upper part) and iron spectral image(lower part) between 553-600nm

Pb II 560.89nm line has been observed at position 1 of Fig 6 but hasn't been observed from Fig 7. Cu lines have been equally observed at position 2 and 3, since they are both copper alloys.

Detection width has been chosen around the selected point in the neighborhood of the wavelength emitted by the element and the program has been designed to output all the elements related to the spectral lines in this area. Correct results have been obtained when the detection width is small.

The detection width must be greater than the inverse degree of dispersion of the given wavelength considering the dispersion curve (Fig 2). With the detection width of $\pm 0.2nm$, the measurement results[21](Fig 6, 7) are 2-Cu570.02nm, 3-Cu1578.21nm and 4-K578.24nm.

The time for obtaining the spectral image of an unknown sample and the desired results is about

60s.

## Conclusions

We have developed a combined system of a spectrometer, a CCD and a computer and applied correlation matching method on the recognition of the digital spectral image, thus giving birth to a new method to rapidly measure the element type, wavelength, relative intensity, sensitivity lines, size of partial image and wavelength area from the spectral image of a given sample.